\documentclass[pra,superscriptaddress,reprint]{revtex4-1}
\usepackage{amsmath}
\usepackage{hyperref}

\usepackage{graphicx,color}

\definecolor{mygreen}{rgb}{0,0.5,0} 
\definecolor{myorange}{rgb}{1.,0.5,0} 
\definecolor{myblue}{rgb}{0,0,0.75} 
\definecolor{mymagenta}{cmyk}{0,1,0,0.12}

\newcommand{\ntext}[1]{{\color{black}#1}}

\newcommand{\ket}[1]{\mid #1 \,\rangle}
\newcommand{\bra}[1]{\langle \, #1 \mid}
\newcommand{\braket}[2]{\langle \, #1 \mid #2 \,\rangle}
\renewcommand{\ket}[1]{| #1 \rangle}
\renewcommand{\bra}[1]{\langle  #1 |}
\renewcommand{\braket}[2]{\langle  #1 | #2 \rangle}

\newcommand{\Eop}[2]{E^{(+)}_{#1}(t_{#2})}
\newcommand{\Eom}[2]{E^{(-)}_{#1}(t_{#2})}
\newcommand{\Eopt}[2]{\tilde{E}^{(+)}_{#1}(t_{#2})}

\newcommand{\PRLsection}[1]{\noindent {\it#1} -}
\newcommand{\psit}{\tilde{\psi}}
\newcommand{\relphase}{\phi_{\rm rel}}

\begin{document}

\title{Interferometric measurement of the biphoton wave function}

\author{Federica A. Beduini}
\email{federica.beduini@icfo.es} 
\address{ICFO-Institut de Ciencies Fotoniques, Av. Carl Friedrich Gauss, 3, 08860 Castelldefels, Barcelona, Spain}

\author{Joanna A. Zieli\'nska}
\address{ICFO-Institut de Ciencies Fotoniques, Av. Carl Friedrich Gauss, 3, 08860 Castelldefels, Barcelona, Spain}

\author{Vito G. Lucivero}
\address{ICFO-Institut de Ciencies Fotoniques, Av. Carl Friedrich Gauss, 3, 08860 Castelldefels, Barcelona, Spain}

\author{Yannick A. de Icaza Astiz}
\address{ICFO-Institut de Ciencies Fotoniques, Av. Carl Friedrich Gauss, 3, 08860 Castelldefels, Barcelona, Spain}

\author{Morgan W. Mitchell}
\address{ICFO-Institut de Ciencies Fotoniques, Av. Carl Friedrich Gauss, 3, 08860 Castelldefels, Barcelona, Spain}
\address{ICREA-Instituci\'{o} Catalana de Recerca i Estudis Avan\c{c}ats, 08015 Barcelona, Spain}



\begin{abstract}
{
Interference between an unknown two-photon state (a ``biphoton'') and the two-photon component of a reference state gives a phase-sensitive arrival-time distribution containing full information about the biphoton temporal wave function.  Using a coherent state as a reference, we observe this interference and reconstruct the wave function of single-mode biphotons from a low-intensity narrowband squeezed vacuum state.  
}
\end{abstract}

\maketitle

\PRLsection{Introduction}
\ntext{Correlated photon pairs, or ``biphotons,'' are a paradigmatic experimental system in quantum technology, with applications in quantum communications \cite{PengPRL2005}, quantum information processing \cite{WaltherN2005}, foundations of physics \cite{GiustinaN2013} and quantum metrology \cite{MitchellN2004}.  In many experiments, the performance of a biphoton source is closely tied to the two-photon wave function (TPWF) that describes the temporal correlations of the photons.  For example, the visibility of Hong-Ou-Mandel interference depends on the TPWF, even when some other degree of freedom, e.g. polarization, is used to encode quantum information \cite{AdamsonPRL2007}.}
\ntext{Measurements of the TPWF are also used to characterize realistic photon pairs sources, allowing the \ntext{diagnosis of} experimental defects, e.g. imperfect poling in the down-conversion crystal~\cite{Kuzucu2008} or dispersion~\cite{ODonnellPRL2009}.}

The TPWF $\psi(t_1,t_2)$ is an intrinsically multi-dimensional object, depending on the two time coordinates $t_1$ and $t_2$ \cite{ValenciaPRL2007}.   Methods to characterize the TPWF include measurement of the joint spectral density \cite{MosleyPRL2008}, measurement of the joint temporal density \cite{Kuzucu2008},  non-classical interference using the Hong-Ou-Mandel effect \cite{sergienko1995,GiovannettiPRA2002,OkamotoPRA2006}, and nonlinear optical processes \cite{DayanPRL2004,PeerPRL2005,ODonnellPRL2009,SensarnPRL2009}.   All of these techniques give partial information about the TPWF.  For example, the joint temporal density gives the  magnitude $|\psi(t_1,t_2)|$, while the joint spectral density gives the  magnitude of Fourier components.

Full measurement of the TPWF requires  a phase-sensitive and tomographic measurement, applied to a continuous range of time values.  Some elements of this approach have been demonstrated:  Quantum state tomography \cite{SmitheyPRL1993} has been widely used to characterize aggregate measures of a quantum state, e.g. the integrated field of a pulse, or the mode describing a single frequency component.  This includes traditional homodyne methods using strong local oscillators  \cite{SmitheyPRL1993} and mesoscopic methods using weak local oscillators plus photon-counting detection \cite{PuentesPRL2009}.  Homodyne characterization of a single photon wave function has also been reported \cite{NeergaardNielsenPRL2006, MorinPRL2013}.

Here we demonstrate full characterization of a two-photon wave function, based on the phenomenon of interference of two-photon amplitudes \cite{TorgersonJOSAB1997,LuPRL2001}.  A similar method is proposed in \cite{RenPRA2012}.  Our approach combines the use of a weak phase reference and photon counting detection as in \cite{PuentesPRL2009} with wave-function detection over an extended time-span as in \cite{NeergaardNielsenPRL2006, MorinPRL2013}, and adds the new elements of time-correlated photon counting, as required by the dimensionality of the TPWF.  We demonstrate the method by reconstructing the TPWF of  single-mode squeezed vacuum from a sub-threshold OPO. An attractive feature of our approach is a very direct data interpretation, without the ill-posed inverse problem typically encountered in tomography.

~

\PRLsection{One- and two-photon wavefunctions}
We use field correlations functions \cite{GlauberPR1963} to characterize  optical quantum states.  For a state $\ket{\lambda}$,  the so-called ``one-photon wave function''  is $\psi_i^{(\lambda)}(t) \equiv \bra{0}\Eop{i}{} \ket{\lambda}$, where $\Eop{i}{}$ is the positive-frequency part of the electric field operator for mode $i$.   Because $\Eop{i}{}$ removes one photon, this represents $\ket{\lambda}$ projected onto the one-photon subspace. Similarly, the ``two-photon wave function'' is \cite{sergienko1995}
\begin{equation}
   \psi_{i,j}^{(\lambda)}(t_1,t_2) \equiv \bra{0}\Eop{i}{1}\Eop{j}{2} \ket{\lambda}.
\end{equation}
As with Schr\"odinger wave functions, neither $\psi_i^{(\lambda)}(t)$ nor $\psi_{i,j}^{(\lambda)}(t_1,t_2)$ is directly observable.  On the other hand, the second-order intensity correlation function
\begin{align}
   g^{(2)}_{ij}(t_1,t_2) &  \propto \bra{\lambda} \Eom{j}{2}\Eom{i}{1}\Eop{i}{1}\Eop{j}{2}\ket{\lambda}
\end{align}
is directly observable in photon pair arrival time distributions.  In the commonly-encountered case that $\ket{\lambda}$ contains no more than two photons, this is proportional to $|\psi_{ij}^{(\lambda)}(t_1,t_2)|^2$. 
The second order correlation function then gives  important but incomplete information about the two-photon wavefunction, as it contains no information on the phase of $\psi_{ij}^{(\lambda)}$, which is a complex function.

~

\begin{figure}[t]
   \centering
   \includegraphics[width=0.98\columnwidth]{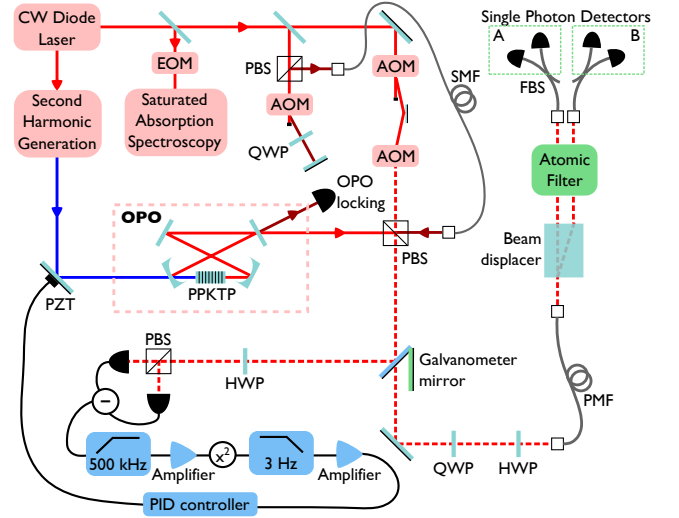}
   \caption{Experimental setup. AOM (EOM): acousto- \mbox{(electro-)} optic Modulator. PBS: polarizing beam splitter. QWP (HWP): quarter- (half-)wave plate. PZT: piezoelectric actuator. SMF: single mode fiber. PMF: polarization maintaining fiber. FBS: fiber beam splitter. }
   \label{fig:setup}
\end{figure}

\PRLsection{Coherent state reference}
We consider a scenario in which $|\lambda \rangle$ occupies one propagating mode ($V$), while a time-independent coherent state $\ket{\alpha}$ occupies an ancilla mode ($H$).  \ntext{
We measure the correlation function 
\begin{equation}
   \psit_{AB}^{(\kappa)} (t_1,t_2)= \bra{0} \Eopt{A}{1} \Eopt{B}{2}\ket{\kappa}\,
\end{equation}
of the global state $\ket{\kappa} = \ket{\lambda}\otimes \ket{\alpha}$ with a polarimeter setup, as \ntext{shown} in Fig.~\ref{fig:setup}: a quarter- and a half-wave plate apply a unitary transformation on the polarization, \ntext{then a beam displacer separates the two polarization components, }so that the field operator associated to detector A(B) \ntext{is} 

\begin{align}
\label{eq:EA}
   \Eopt{A}{} & = \cos\theta \Eopt{H}{}  + e^{ i\phi}\sin \theta \Eopt{V}{} \,\\
\label{eq:EB}
   \Eopt{B}{} & =  e^{- i\phi}\sin \theta \Eopt{H}{}  -  \cos\theta  \Eopt{V}{} \,
\end{align}
}
\ntext{where $\theta$ and $\phi$ are the polar and azimuthal angle in the Bloch sphere, respectively.}

\ntext{
The two-photon wave function of the global state becomes then
\begin{align}
   \psit _{AB}^{(\kappa)} (t_1,t_2)&= e^{-i\phi} \cos\theta \sin \theta \,\psi^{(\alpha)}_{HH}\, \braket{0}{\lambda}\nonumber\\
   &-e^{i\phi} \cos\theta \sin \theta \, \psi^{(\lambda)}_{VV}(t_1,t_2)\, \braket{0}{\alpha}\nonumber\\
   &+\sin^2\theta\,\psi^{(\lambda)}_V(t_1)\,\psi^{(\alpha)}_H(t_2)\nonumber\\
   &-\cos^2\theta\,\psi^{(\alpha)}_H(t_1)\,\psi^{(\lambda)}_V(t_2). 
   \label{eq:psi_kappa}
\end{align}

The last two terms in Eq. (\ref{eq:psi_kappa}) vanish, because $\psi^{(\lambda)}_V(t)   \equiv \bra{0}\Eop{V}{} \ket{\lambda} = 0$ when $\ket{\lambda}$ is squeezed vacuum.  More generally, $\psi^{(\lambda)}_V(t)$ vanishes for any state invariant under $\Eop{V}{} \rightarrow - \Eop{V}{}$ or equivalently $a_V(\omega) \rightarrow a_V(\omega) \exp[i \pi]$.  The symmetry of the down-conversion hamiltonian $H \propto \chi^{(2)} a^\dagger_V a^\dagger_V a_{p} + {\rm h. c.}$, and of dephasing and decoherence processes, guarantees $\psi^{(\lambda)}_V(t) = 0$ in the broad class of experiments using spontaneous, i.e. vacuum-driven, down-conversion.  
}

Taking $\theta = \pi/4$ for simplicity, we can write the measurable second order correlation function as
\begin{align}
   g^{(2)}_{AB (\kappa)} (t_1, t_2) & \propto \left|\gamma e^{-2i\phi}   -\psi_{VV}^{(\lambda)}(t_1, t_2)\right|^2\,,
   \label{eq:g2_inter}
\end{align}
where   \ntext{ $\gamma = \psi_{HH}^{(\alpha)}{\braket{0}{\lambda}} {\braket{0}{\alpha}}^{-1}$}. 
We note that now $g^{(2)}_{AB(\kappa)}$, which is directly measurable, contains information about the phase of $\psi_{VV}^{(\lambda)}(t_1,t_2)$, through interference against $\ket{\alpha}$.  For convenience, we choose the phase origin so that $\alpha$, and thus $\gamma$, is real, and as indicated already $\theta = \pi/4$.   To find $\psi_{VV}^{(\lambda)}$, it is convenient to measure with the azimuthal angle $\phi = k\pi/3$, $k = \{0,1,2\}$, i.e.,  symmetrically placed within the period of $\exp[2 i \phi]$.
We denote the resulting $g^{(2)}$ values as $y_k$.

It is then possible to solve  Eq.~\eqref{eq:g2_inter} 
to obtain the TPWF
\begin{eqnarray}
   \label{eq:psiV}
   \psi_{VV}^{(\lambda)} &=& \frac{\bar{y} -y_0}{2\gamma}+ i \frac{y_1-y_2}{2\sqrt{3}\gamma}\,\\
   \gamma &=& \frac{1}{\sqrt{2}}\sqrt{ \bar{y}+\sqrt{3\bar{y}^2-\frac{2}{3}\left( y_0^2+y_1^2+y_2^2\right)}}\,
   \label{eq:gamma}
\end{eqnarray}
where $\bar{y} \equiv (y_0+y_1+y_2)/3$.  
Note that $\psi_{VV}^{(\lambda)}$, the $y_k$ and $\gamma$ all depend on $(t_1,t_2)$.   
This result is remarkable for its simplicity; the inverse problem to find $  \psi_{VV}^{(\lambda)} $ from the various $g^{(2)}$ measurements gives an analytic solution.

  With the addition of a coherent state, we 
  relate a measurable quantity to the two-photon wavefunction, {recovering both its real and imaginary parts from experimental results.}

~

\PRLsection{Experimental realization}
{To test the technique, we measure the two-photon wave function of weakly-squeezed vacuum from a sub-threshold degenerate optical parametric oscillator (OPO). A continuous-wave diode laser at 794.7~nm generates both the coherent reference beam and, after being amplified and doubled in frequency, a 397.4~nm pump beam for the OPO, described in~\cite{predojevic2008}, which generates a vertically-polarized (V) squeezed vacuum state via spontaneous parametric down-conversion in a periodically poled KTP crystal.  The cavity length is actively stabilized with a Pound-Drever-Hall lock, to keep one longitudinal V mode resonant at the laser frequency.  The locking beam is H polarized, counter-propagating, and shifted in frequency by an acousto-optic modulator (AOM), to match the frequency of an H-polarized mode.  The AOM RF power is chopped and the detectors are electronically gated: coincidence data are acquired only when the locking light is off.  With these measures, the contribution of locking light \ntext{to the accidental coincidences background} is minimised. 

The V-polarized squeezed vacuum is combined with the H-polarized coherent reference at a polarizing beamsplitter to generate a beam with co-propagating squeezed and reference components.  The polarization transformation of Eqs. (\ref{eq:EA}), (\ref{eq:EB}) is implemented with a quarter- and a half-waveplate, and the beam is coupled into a polarization maintaining fiber, with its fast axis aligned to H-polarization when $\theta = \phi = 0$.  
At the fiber output, the two polarization components are separated into parallel beams by a calcite beam displacer and passed through a narrowband (445~MHz) atomic filter~\cite{zielinska2012,ZielinskaARX2014}, in order to isolate the squeezed vacuum and block with high efficiency the hundreds of non-degenerate frequency modes generated by the OPO. The maximum transmission frequency of this filter is located \ntext{at 2.7~GHz to the red} of the center of the rubidium D${}_1$ line,  and the laser frequency is stabilised at this particular frequency by using an integrated electro-optic modulator to add sidebands to the laser prior to the saturated absorption spectroscopy.  
Each filtered beam is then coupled into a single-mode fiber and split with a 50/50 fiber beam splitter to a pair of single-photon counting avalanche photo diodes.  A time-of-flight recorder time-stamps each arrival and correlations are computed on a PC.  

\begin{figure}[bt]
   \centering
   \includegraphics[width=0.98\columnwidth]{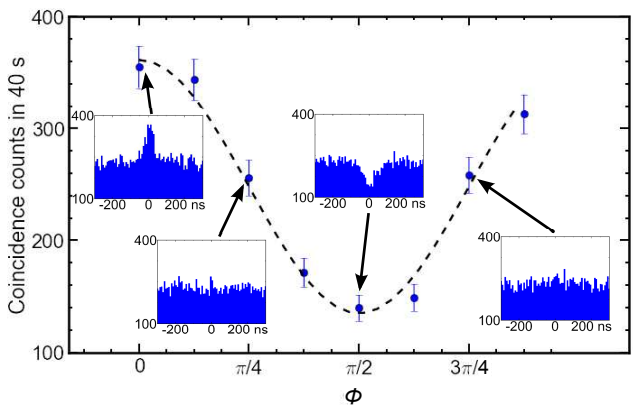}
   \caption{\ntext{\ntext{(Color online). Arrival-time distributions showing interference of two-photon amplitudes.  Main graph shows coincidence rates $g^{(2)}_{AB(\kappa)}(0)$ (circles)} for delay $\tau= 0$ versus analysis phase $\phi$.  These show a sinusoidal behaviour (dashed line, $A + B \cos 2\phi$ fit to the data) revealing two-photon interference as predicted by \eqref{eq:g2_inter}. Insets show $g^{(2)}_{AB(\kappa)}(\tau)$ for the values of $\phi$ indicated with arrows.  These clearly show the passage from constructive interference at $\phi=0$, where a peak is visible, to destructive interference at $\phi=\pi/2$, where a dip appears.  \ntext{Error bars show $\pm 1 \sigma$ \ntext{(standard deviation)} statistical uncertainty. }}   
   }
   \label{fig:visibility}
  
\end{figure}
\begin{figure}[t]
   \centering
   \includegraphics[width=0.98\columnwidth]{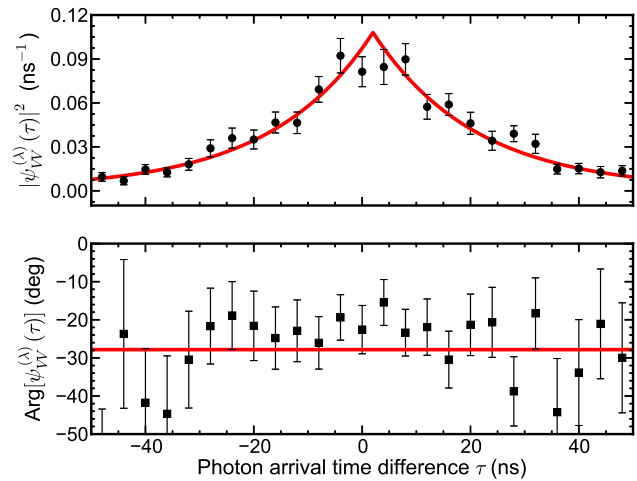}
   \caption{\ntext{(Color online). Squared} amplitude (above) and phase (below) of the reconstructed two-photon wave function for the squeezed vacuum state. \ntext{The solid line shows the predicted, double exponential amplitude describing an ideal squeezed vacuum state from our OPO with an independently-measured 8.1~MHz bandwidth. The amplitude and the horizontal offset were fit to the data.} \ntext{Error bars show $\pm 1\sigma$ statistical uncertainty assuming Poisson statistics and using propagation of error through Eqs. \eqref{eq:psiV} and~\eqref{eq:gamma}.} 
    }  \label{fig:wavefunction}
\end{figure}

A low OPO pump power (1 mW, 0.04\% of threshold) is used so that contributions of more than two photons are negligible.  The coherent reference power is chosen to give a similar rate of two-photon events, for high visibility interference, as seen in Fig.~\ref{fig:visibility}. 
The relative phase $\relphase$ between the coherent and the squeezed beam is stabilized by a quantum noise lock:  One Stokes component is detected with a balanced polarimeter, and the noise power in a 3 Hz bandwidth above 500 kHz is computed analogically using a multiplier circuit.  This signal is fed back by a servo loop to a piezo-electric actuator on a mirror in the pump path, to stabilize the pump phase by a side-of-fringe lock.  A galvanometer mirror is used to switch between the single-photon counting and stabilisation setups at a frequency of $\sim$100 Hz.  The reference beam power is increased during the stabilization part of the cycle, to reach the shot-noise-limited regime optimal for detection of the squeezing and operation of the noise lock. Two cascaded AOMs, whose RF power is chopped synchronously with the galvanometer mirror, modulate the coherent reference beam power, so that it has high power when the light is entering the stabilisation setup and low power when the photon counting part is active. The system can maintain a fixed  $\relphase$  over several hours.}

\PRLsection{Results}  
\ntext{As our light source is continuous-wave, the light statistics are stationary: }
the correlations and wave function depend only on the photon arrival-time difference $\tau=t_1-t_2$.  We compute the experimental $g^{(2)}_{AB(\kappa)}(\tau)$ from coincidences between detector groups A and B in Fig.~\ref{fig:setup}, with a 4~ns coincidence window, a compromise between temporal resolution and statistical significance.  

As shown in Fig.~\ref{fig:visibility}, we  observe both constructive and destructive interference, e.g. at $\phi=0$ and $\phi=\pi/2$, respectively.
The observation of a dip in the correlation function is especially interesting, because it clearly signals destructive interference of two-photon amplitudes from the coherent and the squeezed vacuum states.
The interference visibility is limited by accidental coincidence counts, which are mainly due to the residual OPO locking beam and to non-degenerate modes passing through the filter~\cite{ZielinskaARX2014}. \ntext{However, these do not affect the wavefunction reconstruction: the accidentals add a  term independent from $\tau$ to the $g^{(2)}$, which is canceled by the subtractions in Eq.~\eqref{eq:psiV}.}

We next collect $g^{(2)}_{AB(\kappa)}(\tau)$ data for $\phi = 0,\pi/3,2\pi/3 $ and use Eqs.~\eqref{eq:psiV} and~\eqref{eq:gamma}  to reconstruct 
$\psi_{VV}^{(\lambda)}(\tau)$, shown in Fig.~\ref{fig:wavefunction}.  The reconstruction is direct: $\psi_{VV}^{(\lambda)}$ at a given $\tau$ depends only on coincidence events at that value of $\tau$.  The results are consistent with a double-exponential amplitude with 26~ns full-width at half-maximum (FWHM), as expected for a squeezed vacuum state from an OPO with the 8.1~MHz FWHM bandwidth independently-measured on our system.  Fig.~\ref{fig:wavefunction} also shows a constant but nonzero phase of the wave function.  A constant phase is expected for an ideal OPO, while a phase defect could signal cavity or crystal imperfections \cite{Kuzucu2008,ODonnellPRL2009}.  The phase offset is tunable via the side-of-fringe lock that sets the relative phase of the squeezed vacuum and reference, and is another indication of interference at the two-photon level.

~

\PRLsection{Conclusion}
\ntext{We have demonstrated  complete measurement of the complex temporal wave function of biphotons using interference of the two-photon amplitude against a reference.  The interference gives a phase-sensitive arrival-time distribution, from which we reconstruct the biphoton wave function. 
  In contrast to most tomographic procedures~\ntext{\cite{AdamsonPRL2007,SmitheyPRL1993}}, only three measurement settings are required to find the real and imaginary parts of the wave function, as well as the strength of the reference state.  The inverse problem is thus neither over-determined nor under-determined, and can be solved analytically.  We analyze the output of a narrow-band, atom-resonant OPO operating at 795 nm, and find a biphoton wave-function consistent with squeezed-vacuum biphotons from an ideal OPO with our measured line-width.  

The technique shows clearly the interference of two-photon amplitudes from distinct sources, and may be
useful for detecting and correcting errors in quantum light sources for quantum information processing \cite{WolfgrammOE2008}, quantum communications \cite{FeketePRL2013}, and quantum metrology  \cite{WolfgrammNPhot2013}. }

~

\PRLsection{Acknowledgements}
\ntext{ We thank F. Wolfgramm and F. Martin Ciurana for helpful discussions.} This work was supported by the Spanish MINECO project MAGO (Ref.
FIS2011-23520), European Research Council project AQUMET and by Fundaci\'{o} Privada CELLEX. J.Z. was supported by the FI-DGR PhD-fellowship program of the Generalitat of Catalonia. \ntext{Y. A. de I. A. was supported by the scholarship BES-2009-017461, under project FIS2007-60179.}
\bibliographystyle{apsrev4-1no-url}
\bibliography{2ph_wavefunction,BigBib130123}

\end{document}